# A combined dose and microdosimetric modeling framework incorporating volume effects correlates with tissue sparing in proton minibeam radiotherapy

Bordieri G.[1,2], Battestini M.[1,2], Lattanzi G.[1,2], Romano, F.[3], Scifoni E.[2], Missiaggia M.[2,4,*], Cordoni F.G.[2,5]

1 Department of Physics, University of Trento, via Sommarive 14, Trento, 38123, Italy

2 Trento Institute for Fundamental Physics and Application (TIFPA), via Sommarive 15, Trento, 38123, Italy

3 Istituto Nazionale Fisica Nucleare (INFN), Sezione di Catania, Catania, Italy

4 The Department of Physics & Astronomy at Louisiana State University (LSU), 202 Nicholson Hall, Baton Rouge, LA 70803, US

5 Department of Civil, Environmental and Mechanical Engineering, University of Trento, via Mesiano 77, Trento, 38123, Italy

*Corresponding author

E-mail: giulio.bordieri@unitn.it, marco.battestini@unitn.it, emanuele.scifoni@tifpa.infn.it, gianluca.lattanzi@unitn.it, francesco.romano@ct.infn.it, emanuele.scifoni@tifpa.infn.it, mmissi2@lsu.edu, francesco.cordoni@unitn.it

# Abstract

***Purpose***

Proton minibeam (pMB) radiotherapy, delivers highly heterogeneous dose distributions alternating high-dose peaks and low-dose valleys. This aims to widen the therapeutic window by improving normal tissue sparing while maintaining the same or even better tumour control. The performance of pMB strongly depends on the collimator design and physical parameters. To better understand the physical and radiobiological drivers of this enhanced therapeutic window, we perform a detailed microdosimetric characterization of proton minibeams and assess their impact.

***Methods***

We characterize radiation quality with microdosimetry through Monte Carlo simulations. Then we extend the Generalized Stochastic Microdosimetric Model ($GSM^2$) to predict the normal tissue complication probability (NTCP) at different depths in water, 1cm, 2cm, and 4cm, for 100MeV proton minibeams realized with varying configurations of collimator. Results are compared with conventional homogeneous field (HF) irradiation after dose normalization to the tumor. The developed model is applied by considering tissues as divided into several functional subunits, connected by introducing a seriality parameter.

***Results***

Microdosimetric characterization of proton minibeam irradiation shows differences between peak and valley regions in shaping lineal energy spectra, especially at low depth, while radiation quality uniforms progressively getting closer to the target region

(tumor). NTCP calculations results suggest an increased sparing effect for pMB over conventional HF. A strong dependence is found on the peak-to-valley dose ratio (PVDR), and on the seriality parameter. Predictions indicate substantial sparing from pMB, especially for PVDR>15, including relatively serial organs with seriality around 0.7. All results are consistent with the general trends reported in experimental studies.

*Conclusion*

This integrated dose–microdosimetric–biological framework elucidates how spatial fractionation, radiation quality, and organ architecture collectively shape tissue sparing in pMB. The findings identify conditions under which pMB may offer NTCP reduction, highlighting the importance of incorporating microdosimetry and tissue seriality in future optimization and clinical translation efforts.

# Introduction

Spatially fractionated radiotherapy (SFRT) encompasses techniques that deliver spatially heterogeneous dose distributions, typically high-dose peaks interspersed with low-dose valleys, within a tumour volume or normal tissue (NT) region. Among SFRT modalities, proton minibeam (pMB) has recently attracted attention as a good compromise between clinical feasibility and effectiveness. pMB delivers proton beams with widths on the order of a millimeter, and it has been shown to enhance the therapeutic window by sparing normal tissue while maintaining, or even improving, tumor control compared to conventional radiotherapy[1,2,3,4].

The main advantage of this irradiation technique is the treatment of deep-seated tumors by enhancing NT sparing and exploiting the physical effect of multiple Coulomb scattering of minibeams to not deteriorate tumor control. On the other hand, some experimental studies have resulted in observing a direct higher killing efficiency on superficial tumors treated with a spatially fractionated radiation dose[5,6], most likely linked to the trigger of an enhanced immunotherapeutic response[7,8]. The enlarging of the therapeutic window due to pMB has been investigated in both experimental and theoretical results with foundations on Monte Carlo (MC) simulations[9,10,11,12,13,14]. However, to date, a clear mechanistic understanding of the emergence of this double effect in pMB has yet to be achieved by the community. Several hypotheses have been proposed in recent years to explain the observed benefits of pMB as compared to homogeneous field (HF). In minibeam irradiation of deep-seated tumors, experimental evidence indicates that normal-tissue sparing primarily arises from the substantially reduced irradiated volume within organs at risk, thereby enhancing their capacity for biological repair. Nevertheless, the elevated peak doses relative to a HF introduce a complex radiobiological response, preventing the direct application of standard clinical NTCP models.

To achieve this spatial fractionation of the dose, specifically designed collimators are employed to generate well-defined minibeam arrays. Different thickness values have been explored in different studies, both experimentally and with MC simulations[15,16,17,18,19] to evaluate the impact of this parameter on the pMB distribution. Experimental evidence suggests that the fundamental parameter to consider is the center-to-center (CTC) distance among the peaks, in combination with the slit width, which has an impact on the physical separation of the minibeams and consequently on the homogeneity of the dose deposited on deep-seated tumors. Other studies report peak-to-valley dose ratio (PVDR), defined as the ratio of the dose in the peak regions to that in the valley regions, or recently also the valley dose has been correlated to normal tissue sparring[20,21,22].

Since valley regions are characterized by fewer particles that scatter from the peak regions, pMB not only creates inhomogeneous dose patterns but also inhomogeneous radiation quality distributions, i.e., particle type and energy spectrum. However, radiation quality, as microdosimetry, has not been investigated extensively by the community in pMB, with some results only for carbon ions[23].

This work characterizes radiation quality with MC simulated microdosimetry at the center of the peak and valley, with the final goal of providing an in-silico radiobiological modelling study of pMB particle therapy. This information, alongside dose, is used to calculate the normal tissue complication probability (NTCP) through microdosimetry-based radiobiological modeling[24,25,26,27,28]. This interpretation provides the conceptual basis for our modelling strategy, which starts from microdosimetric spectra to calculate survival probabilities for single cells and cell populations using the Generalized Stochastic Microdosimetric Model ($GSM^2$), explicitly incorporating the seriality parameter that characterizes different biological tissues to extend the damage prediction results to the organ level.

Building on this foundation, we formulate a general NTCP model derived from the critical-volume (CV) model[29]. The CV model is an NTCP model used in radiotherapy to describe how damage to a specific minimum fraction of functional subunits within an organ leads to clinical complications. The CV assumes that an organ is represented as an ensemble of functional subunits (FSUs)[30] whose collective response determines the macroscopic clinical outcome. This assumption is consistent with standard clinical practice and reflects a widely used setting in NTCP modelling, which is based on the dependence of tissue radiation response on the irradiated volume, better known as volume effects[31,32]. Within this framework, we employ our radiobiological model to calculate responses at both the individual FSU and organ levels, enabling us to describe broadly heterogeneous irradiation conditions. Our approach allows us to treat heterogeneity not only in terms of absorbed dose, but also in terms of radiation quality quantified through microdosimetry.

Based on this formulation, we systematically investigate how different pMB configurations lead to distinct biological outcomes depending on the assumed degree of organ seriality. Our findings highlight how the interplay between volume effects and a detailed analysis of dose and radiation quality can partially account for the observed biological advantages of pMB.

## Materials and Methods

### Monte Carlo simulations and microdosimetric analysis

We consider four different pMB irradiation patterns for 100 MeV protons compared to the conventional homogeneous field, and we simulate the dose deposition within a 10cm-thick water phantom with the Geant4 MC simulation toolkit (version 11.2.0). Configurations are chosen to span from cases in which the peak width is much lower than the valley width, up to cases with equal width between peak and valley, which is investigated for both narrow beams and 1mm-size peaks. A uniform proton beam consisting of $10^8$ particles is delivered towards the phantom, and the four pMB configurations are realized by changing the physical properties of the collimator, which is made of tungsten with a 5cm thickness to ensure stopping of all particles in pMB valleys.

The four pMB configurations have the following parameters in terms of peak width-CTC in mm (associated configuration number): 0.4-3.2 (1), 0.6-1.2 (2), 1-2 (3), 1-3.8 (4). Furthermore, results have been evaluated at three different values of depth in NT: 1cm, 2cm, and 4cm.

A microdosimetric investigation is carried out to characterize the radiation field and assess potential differences in radiation quality at different depths among the pMB peak center, the pMB valley center, and the HF condition. Microdosimetric spectra are scored within a 1µm-radius water sphere as a sensitive volume and used as an input to calculate the survival probability.

We considered the microdosimetric frequency-mean lineal energy ($y_F$) and the dose-mean lineal energy ($y_D$), defined as follows[33]

$$\overline{y_F} = \int_0^\infty y f(y) dy$$
$$\overline{y_D} = \frac{1}{\overline{y_F}} \int_0^\infty y^2 f(y) dy,$$

where $f(y)$ is the lineal energy probability distribution.

### NTCP calculations

We extend the CV model[34] to include cell inactivation by means of GSM$^2$ [35][25][36][28]. The complete derivation is reported in the Appendix. Following recent papers [37][38], we can use GSM$^2$ to calculate the probability that a cell population survives radiation starting from a microdosimetric characterization of the radiation field. We thus assume that an organ is subdivided into FSUs[30] and that preservation of fewer than 100 x s% of FSU leads to organ inactivation, where *s* is the seriality parameter. The calculations and analysis are performed at different depths by considering different NT layers, whose geometry is divided into $N_{FSU}$ = 40 cubic FSUs with a side of 0.5mm, and in each FSU, we estimate around 5000 clonogens, which is consistent with the typical cell density of $10^4 - 10^5$ cells/mm$^3$ [39]. For each clonogen, the survival probability S is calculated considering the corresponding microdosimetric spectrum depending on the position within the pMB irradiation pattern. The FSU overall probability of death is defined as the probability that all the cells die after irradiation, and thus it is defined as the product of 1-S from all cells.

We consider a seriality parameters s in the range [0,1], where s defines the fraction of FSUs that must remain functional for the organ to survive. In other words, the organ is considered intact if at least 100 x s% of its FSUs are preserved. Therefore, s=0 means a fully parallel organ, so all FSUs need to be dead to kill the whole organ, whereas s=1 is a fully serial organ where a single FSU inactivation leads to organ inactivation.

To estimate the NTCP, we develop a tailor-made MC framework as detailed in the pseudo code in the Appendix, designed to account for the fact that each cell in the FSU may receive a different dose and radiation quality, and therefore exhibit a distinct survival probability.

We perform this calculation for the HF field scenario and all four pMB irradiation configurations, for three different depths: 1cm, 2cm, and 4cm. The doses are always normalized to guarantee the same uniform dose to the tumor, considered to be at 7.5 cm in depth. This automatically guarantees that all pMB and HF configuration has the same tumor control, allowing for estimating the advantage in terms of NTCP with unvaried TCP.

From the NTCP results, we extract the dose corresponding to 50% probability, named $D_{50}$. This value refers to the integral dose of all FSU so that the organ has 50% probability of developing complication. Then, we compare pMB and HF, in terms of the ratio of $D_{50}$ values. We define what we call Minibeam Dose Ratio at 50% ($MDR_{50}$) as the ratio between $D_{50}$ for HF and $D_{50}$ for pMB:

$$MDR_{50} = (D_{HF}/D_{PMB})_{50}$$

This results in a value below 1 if the pMB configuration presents an advantage compared to HF. On the other hand, if the value is higher than 1, this means that the radiation-induced biological effect is higher for the spatially fractionated irradiation. Since the $MDR_{50}$ depends on the relatively seriality of an organ, we will investigate this parameters as a function of s.

To evaluate the potential improvement in terms of NTCP from pMB, we define a second estimator that we call Minibeam Benefit Integral (MBI). When we represent the $MDR_{50}$ against the seriality parameter, the MBI is defined as the area between the $MDR_{50}$ curve and the horizontal threshold at 1. This can be mathematically described as

$$MBI = \int_0^{\bar{a} \mid MDR_{50}(\bar{a}) = 0} (1 - MDR_{50}(s))\, ds,$$

where s is the seriality parameter.

The higher the MBI, the higher the impact of the specific pMB configuration on reducing the NTCP compared to conventional irradiation at a specific depth in water. The MBI has the advantage of accounting for the seriality parameter below which HF becomes less advantageous and pMB becomes more favourable. Therefore, the higher the MBI, the more serial an organ can be and still benefit from pMB.

The detailed framework of the computation is reported in **Appendix A**.

The parameters of the model are set to obtain at the entrance (1cm) a $D_{50}$ value between 22Gy and 30Gy, depending on the organ seriality, for conventional homogeneous irradiation, which is comparable with experimental results [40].

# Results

**Figure 1** shows the 2D dose maps, where the depth inside the water phantom is represented in the horizontal axis against the longitudinal dose on the vertical axis, perpendicularly to the collimator slits.

**Figure 2** collects the longitudinal projection of the dose distribution to focus on the pMB patterns for the different scenarios. The plot shows the dose normalized to the mean value at 7.5 cm depth, where the Planning Target Volume (PTV) is placed. This is marked with the red line, while in blue we consider the dose at the entrance 1cm deep inside the water phantom.

As we understand also from **Figure 2**, the PVDR can be estimated by varying the depth. We show all these data in **Figure 3**, where different colors apply to different pMB configurations. On the x-axis is the depth from 1cm up to the considered tumor position of 7.5 cm, while the PVDR is on the y-axis. Configuration 1 is represented with blue dots and reaches the maximum PVDR value at the lowest evaluated depth (1cm) set at around 200, configuration 2 is represented in red, while configurations 3 and 4 are shown in green and yellow, respectively.

The radiation quality analysis performed with the calculation of the $\overline{y_D}$ and $\overline{y_F}$ microdosimetric values is reported in **Figure 4**. In panels (a) and (b), results are compared for the center of the peak and of the valley for pMB and HF. In both cases, results for the microdosimetric values on the y-axis (logarithmic scale) are represented against depth in centimeters on the x-axis. Results for HF configuration are in blue, while we use red and green dots to plot the center of minibeam peaks and valleys, respectively.

The last panel leaves the focus only on pMB and represents the ratio for the two microdosimetric quantities between peak and valley. Blue dots report $\overline{y_D}$, while red $\overline{y_F}$.

The NTCP results are shown in **Figure 5** by plotting the $MDR_{50}$ against the seriality parameter on the horizontal axis. Results are shown for every pMB configuration by color and for different depths in water in the different panels. The color code is the same for all three scenarios, as for **Figure 3**: configuration 1 in blue, 2 in red, 3 in green, and 4 in yellow dots. Panel a) shows the results for 1cm in depth, and we note that the seriality threshold around which pMB are effective in enlarging the therapeutic window is between 0.5 and 0.7, depending on the configuration, where $MDR_{50}$ gets lower than 1. However, for high levels of seriality, $MDR_{50}$ can exceed 3, as for configuration 1. Panel b) and c), respectively representing results at 2cm and 4cm, show a progressive flattening of the $MDR_{50}$ around 1.

**Figure 6** summarizes the MBI for each pMB configuration as defined in the previous section. Here, we can visualize the results for the used metrics in the different studied cases and for the analyzed depths against the PVDR in a semi-logarithmic plot on the horizontal axis. The same colors are used to indicate the different pMB configurations, while different shapes correspond to different depths in the water phantom: 1cm

(square), 2cm (circle), and 4cm (triangle). The monotone increasing trend with the PVDR goes from almost 0 up to over 40%.

# Discussion

From MC dose simulations in **Figure 1** and **Figure 2**, it is evident that the typical pMB irradiation pattern for all four analyzed configurations is a progressive homogeneity while the beam traverses the medium, as known from scattering processes. Moreover, we find that the higher the CTC compared to the peak width, the higher the local entrance dose compared to the dose on the PTV, and the lower the level of homogeneity we reach at the tumor depth. This is a key point of our investigation. This dose homogeneity within the tumor ensures that all pMB and HF configurations achieve the same tumor control, allowing us to evaluate differences in NTCP while keeping TCP constant. Therefore, all NTCP analyses can be considered iso-TCP, meaning the reported values represent a pure gain or loss in terms of NTCP. It is, however, worth stressing that pMB is known to also improve tumor control. This study is left out of the current work, being the primary focus on the NT sparing. Therefore, the therapeutic window can be further widened by better tumor control.

These results are clear in **Figure 2**, where configuration 1 presents local dose values on the peaks that are 1.5 times higher than the average dose at the tumor volume. So, by looking at the spatial distribution of each pMB pattern, we can associate a deviation from the HF case, and this has consequences on the NTCP calculation. The main difference from one pMB configuration to another is summarized in **Figure 3**, where the PVDR is represented against depth in the water phantom. As the CTC increases, the PVDR is higher, especially if the CTC-peak width ratio is very large. So, the higher PVDR results are obtained for pMB configuration 1, while the lowest values are from configuration 2. The field gets more homogeneous with increasing depth, and this is reflected in the decrease of PVDR closer to 1.

Radiation quality is investigated extensively in **Figure 4**, where we show the main microdosimetric estimators, namely $\overline{y_D}$ and $\overline{y_F}$, for pMB peaks and valleys and for HF. First, both trends along the depth in water are found to be in line with recent experimental microdosimetric results [41]. Slightly higher values than experimental values reported are to be imputed to the scoring volumes to be a 1µm water sphere. This choice reflects a radiobiological modeling perspective, using the ideal sensitive volume defined as the microdosimetric scorer. Moreover, we find that the analyzed quantities for pMB peaks and the broad beam in HF do not show significant differences, which arises from the very similar energy spectrum of the particles in HF and traversing the slits of an pMB collimator.

We show more interesting results regarding the difference between the pMB peak center and the valley center. For both $\overline{y_D}$ and $\overline{y_F}$ we record a significant difference in comparing values from peaks and valleys up to 4cm in depth, after which the homogeneity of the field is also reflected in radiation quality in terms of microdosimetry. However, as it is shown in panel c) of **Figure 4** from the ratio of the microdosimetric estimators from peak and valley, we observe an opposite behavior between $\overline{y_D}$ and $\overline{y_F}$. For homogeneous fileds $\overline{y_F}$ is higher, and $\overline{y_D}$ is lower in valleys than in peaks. This comes directly from the energy selection operated with the collimation procedure. In fact, particles reaching the center of the valley region at very low depths due to scattering processes are generally lower energetic than the primaries traversing the medium in the peaks, but still, they need to have sufficient momentum to get to the center of the valley. Comparing valleys to peaks, this results in microdosimetric spectra with a main peak shifted to the higher lineal energy values, but without a very high lineal energy component. For this reason, the resulting values of $\overline{y_F}$ are higher in pMB valleys, while $\overline{y_D}$ values are higher in pMB peaks, where the high lineal energy tail has a stronger weight. These complex inhomogeneities in radiation field quality must be weighted with the specific local dose deposition and require a radiobiological model based on the treatment of the full microdosimetric spectrum to compute cell survival probability and NTCP for pMB. In this sense, the in-silico results presented show that MB fields extreme spatial modulation not only reshapes dose deposition but fundamentally alters the underlying radiation-quality landscape, making them qualitatively different from any other radiotherapy modality.

Once the NTCP is calculated for all the analyzed values of depth in the water phantom, namely 1cm, 2cm, and 4cm, we focus on the $MDR_{50}$ between pMB and HF for each configuration. Results in **Figure 5** show this quantitative measure of the effectiveness of pMB against the seriality parameter associated with the NT, and all four pMB configurations are represented with different colors. We recall that, if the $MDR_{50}$ is lower than 1, this means pMB presents an advantage over HF conventional treatment. On the other hand, the higher the $MDR_{50}$ compared to 1, the higher is the damage caused to NT in pMB conditions.

The general trend for all pMB configurations is a monotonically increasing trend with the seriality parameter, and this is due to the typical dose distribution observed in **Figure 1** and **Figure 2** and coherent with organ seriality. Large dose inhomogeneities favor parallel organ in sparing and on contrary serial organs are inactivated also by high enough local dose deposition. Each pMB pattern has a slightly different seriality parameter threshold that triggers the passing from an advantage to a disadvantage situation, which ranges from around 0.5 in the case of configuration 2 to 0.7 in the case of configuration 1. This seriality threshold mainly depends on the valley width in the specific configuration, while the quantitative deviation from the conventional HF introduces a dependence also on the peak width. This can be observed by comparing configurations 1 and 4, where the same valley width sets the seriality threshold for pMB benefit around 0.7, while the larger peak width for configuration 4 induces an $MDR_{50}$ that is overall closer to 1, suggesting the importance of CTC. This shows the non-conventional effects produced by pMB, where the enlarging of the therapeutic window is also present for relatively serial organs at around 0.7, pointing to a potential benefit of pMB across a wide range of organs.

In the three panels, the depth in water changes, and the global trend is an alignment towards $MDR_{50}$ equal to 1. This confirms the homogeneity of the field, because if $D_{50}$ = 1 for all seriality levels implies that there are no difference between pMB and HF.

A quantitative evaluation of the pMB advantage in terms of MBI is represented against the PVDR **Figure 6**. We recall that this metric was introduced to simultaneously account for two aspects: the increase in dose that a pMB can deliver without altering the biological effect compared to HF, and the ability to include information on how serial an organ can be while still benefiting from a pMB. The higher the PVDR, the higher the MBI, which goes up to over 40% for configuration 1 at 1cm depth. This follows a logarithmic trend by aligning the information on all different configurations for all three values of depth in water. We can note again that configurations 1 and 4 show a higher NTCP advantage compared to the other two pMB configurations at the same depth, with MBI values over 15% for all depth values inside the tissue. In general, for all configurations, higher depth results in a lower MBI as field homogeneity dominates, reaching values close to 0 for configurations 2 and 3. This confirms that even relatively serial organs can benefit from pMB, and that this advantage can be systematic and significant across a broad range of seriality conditions.

Overall, a complex pattern emerges, with multiple parameters influencing the radiobiological effect and supporting the extensive experimental evidence. Our results highlight that both peak and valley doses contribute differently to the overall outcome, and no universal values exist; instead, they should be calibrated according to specific organs and their response characteristics. Finally, PVDR shows a clear correlation with the general effect, with higher PVDR values providing greater benefits, including for relatively serial organs.

We emphasize, however, that our explanation of pMB remains partial. The current model does not include the full biological mechanisms, nor does it capture immunological and bystander-type effects known to contribute to the clinical response in pMB[42,6,7,8]. These limitations indicate that the present radiobiological description represents only one component of a more complex, multi-scale picture. Nonetheless, the results suggest that integrating microdosimetric information with organ-level structural parameters provides valuable insight into the physico-biological mechanisms governing NT response under highly heterogeneous irradiation fields, and constitutes an important step toward overcoming current barriers to the consistent clinical adoption of pMB.

# Conclusions

This work provides a comprehensive in-silico investigation of pMB, integrating MC–based microdosimetric characterization with a microdosimetry-driven NTCP modeling approach with GSM$^2$. By analyzing several minibeam configurations across multiple depths in water, we show that spatial fractionation induces pronounced differences in both dose deposition and radiation quality between peaks and valleys, particularly in the entrance region. These variations progressively diminish with depth as the field approaches homogeneity near the target.

NTCP predictions highlight that minibeam irradiation can substantially enhance NT sparing relative to homogeneous fields, and that this benefit depends on both the physical beam-shaping parameters, most notably the PVDR and CTC, and the biological properties of the tissue, considered by the seriality parameter. While highly parallel tissues consistently benefit from pMB delivery, even organs with a relatively high degree of seriality (e.g., around 0.7) still exhibit a clear and substantial sparing effect, with only extremely serial configurations approaching conditions in which the advantage becomes limited.

Overall, our findings reinforce the relevance of coupling detailed microdosimetry with biological modeling to understand the complex interplay between dose, radiation quality, and tissue architecture in pMB.

# Fundings

This work has been partially supported by the INFN CSN5 project MIRO and PRESTO financed by PIANOFORTE.

# Data Sharing Statement

All data generated and analyzed during this study are included in this published article (and its supplementary information files).

# Appendix A

**Standard formulation for GSM²**

For GSM², the cell survival for a cell domain is calculated as

$$S_n(z_n) = \left( p_0^X(0|z_n)\, p_0^Y(0|z_n) + \sum_{x_0=1}^{\infty} p_0^X(x_0|z)\, p_0^Y(x_0|z_n) C(x_0) \right)^{N_d},$$

Where $p_0^X(0|z_n)$ and $p_0^Y(0|z_n)$ represent the probability that the domain suffers $x_0$ sub-lethal lesions and 0 lethal lesions, respectively,

$$p_0^X(x) = \int_0^\infty p_z^X(x|\kappa z)\, f(z|z_n)\, dz$$

$$p_0^Y(y) = \int_0^\infty p_z^Y(y|\lambda z)\, f(z|z_n)\, dz$$

And $C(x_0)$ are weighting terms representing the probability that $x_0$ sub–lethal lesions are repaired, and thus the domain survives:

$$C(x_0) := \frac{\rho(x+1)\rho(x+2)\rho(x+3)\ldots\rho(x_0)}{(\gamma(1))\ldots(\gamma(x_0))}$$

Also, we set $\rho(x_0) := (x_0)r$ and $\gamma(x_0) := \big((a+r)x_0 + bx_0(x_0-1)\big)$. We can sample $z_n$ from the multievent microdosimetric distribution to obtain the probability that a cell that received dose $z_n$ survives $S_n(z_n)$.

**Survival probability for a FSU**

We can thus calculate the probability that an FSU is inactivated, $P^{FSU}$ as the probability that all clonogens in the FSU will die. In a FSU composed by N clonogens, this is given by

$$P^{FSU} = \prod_{j=1}^{N} (1 - S(z_n^j))$$

**Whole organ effect**

We model an organ as composed of M FSUs and introduce a seriality parameter $s \in [0,1]$, which characterizes the organ's architecture:

- s = 0: fully parallel organ (organ fails only if all FSUs are inactivated),
- s = 1: fully serial organ (organ fails if even one FSU is inactivated).

The parameter m defines the fraction of FSUs that must remain functional for the organ to survive. In other words, the organ is considered intact if at least 100 x s% of its FSUs are preserved. Consequently, the NTCP corresponds to the probability that more than 100 x s% of FSUs are inactivated. Because the model accounts for complete dose and radiation quality inhomogeneity, NTCP cannot be evaluated analytically and must be computed numerically. The following pseudo-code outlines the procedure used to estimate NTCP.

## Monte Carlo Simulation for NTCP Calculation

The following pseudocode outlines our custom Monte Carlo method for computing the NTCP.

```
1. Cell-Level Survival
   For each Functional Subunit (FSU), containing N clonogenic cells
    :
       For each clonogen i in {1, ..., N}:
           - Sample local energy deposition z_i
             from the microdosimetric distribution.
           - Compute cell survival probability S_i = S(z_i).

2. FSU-Level Survival
   For each of the M FSUs:
       - Compute FSU survival probability P_FSU
         as the product of each cell's probability of dying.
       - Draw U ~ Uniform(0, 1).
       - If U < P_FSU:
             Mark FSU as inactivated.

3. Organ-Level Assessment
       Let W = (number of inactivated FSUs) / M.
       If W > m:            # m = organ seriality parameter
           NTCP_MC = 1
       Else:
           NTCP_MC = 0

4. Simulation Loop
       Repeat Steps 13 for K = 10^6 Monte Carlo iterations.
       Final estimate:
           NTCP = (sum of NTCP_MC) / K
```

| Configuration # | Peak width [mm] | CTC [mm] |
|---|---|---|
| 1 | 0.4 | 3.2 |
| 2 | 0.6 | 1.2 |
| 3 | 1 | 2 |
| 4 | 1 | 3.8 |

*Table A.1: Main physical parameters for the different MB configurations considered in this study*

# Figures

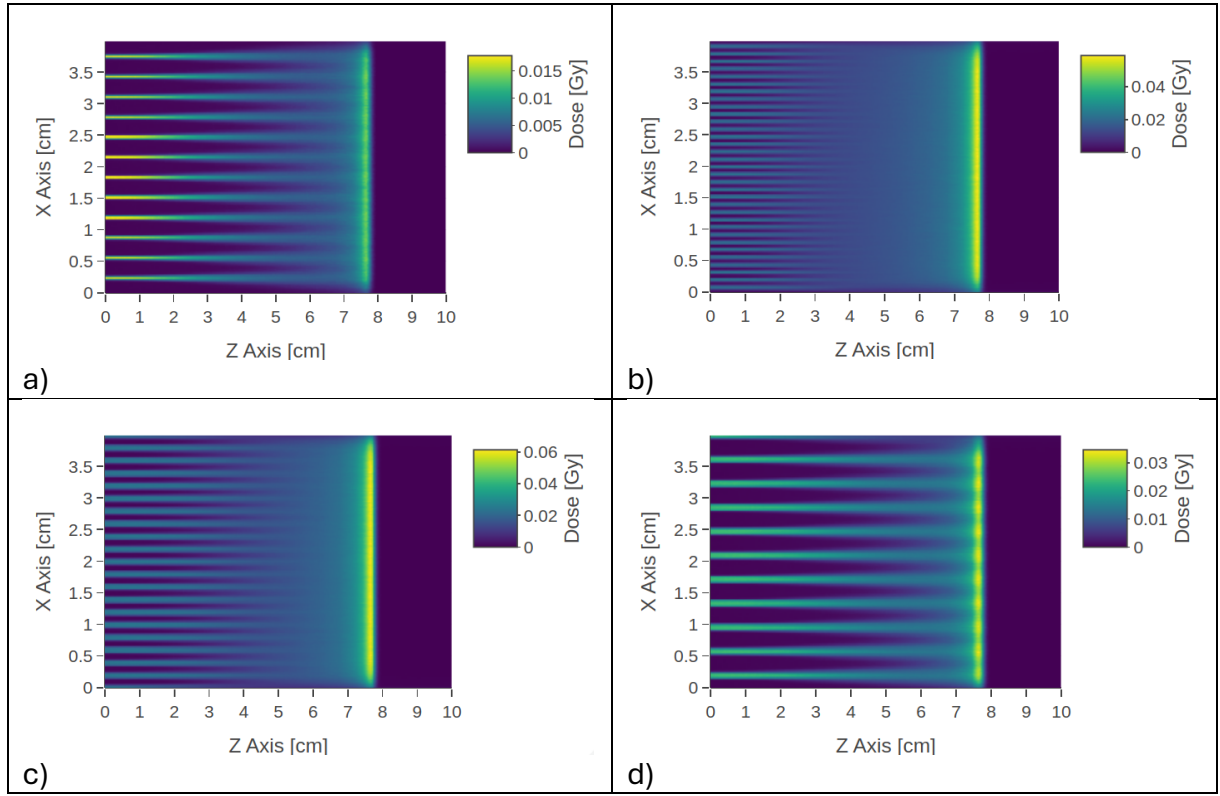

**Figure 1**: 2D dose distribution maps for all four analyzed MB configurations. Configurations in terms of peak width - CTC in mm are: 0.4-3.2 (1) in panel a), 0.6-1.2 (2) in panel b), 1-2 (3) in panel c), and 1-3.8 (4) in panel d).

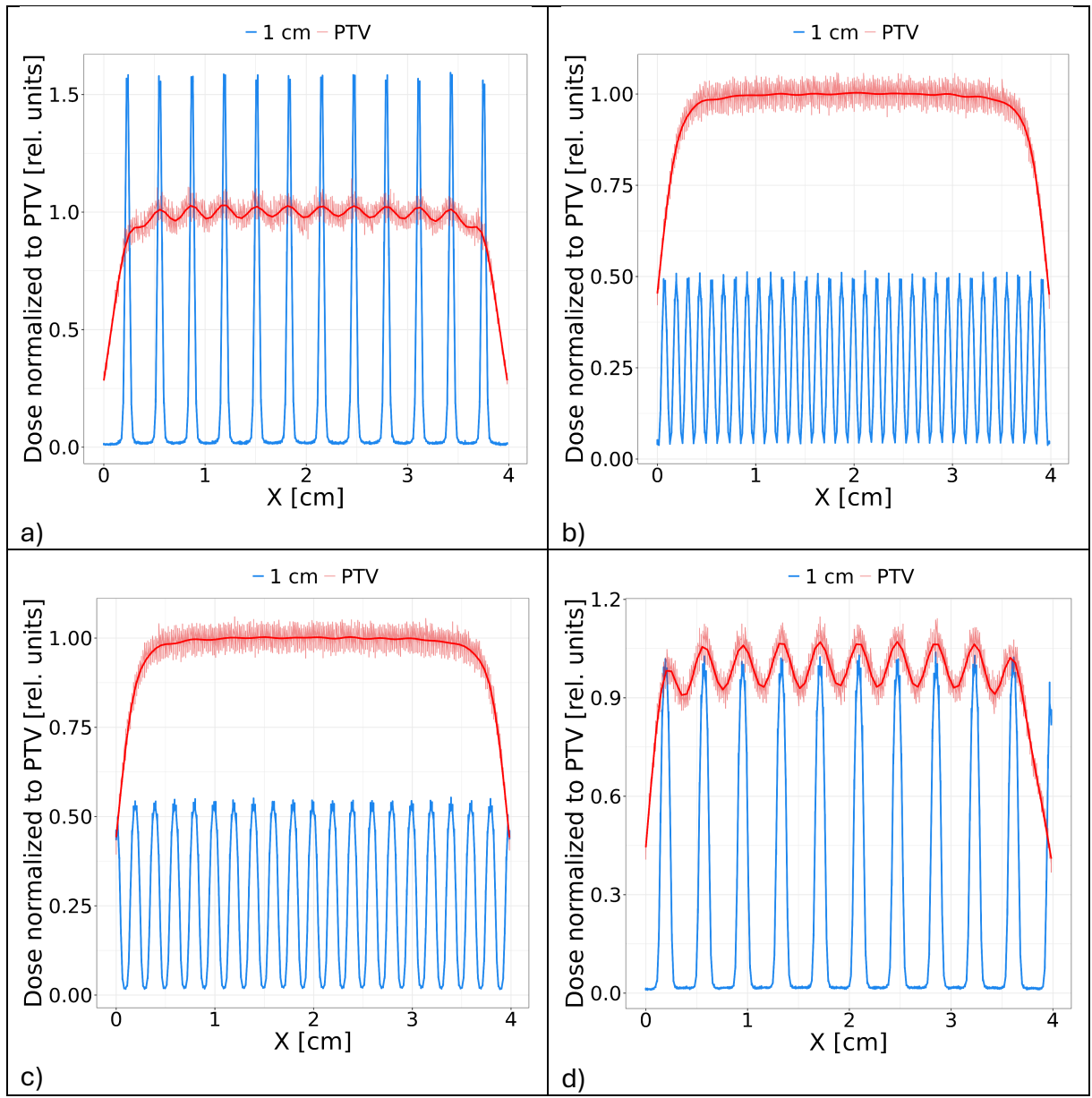

**Figure 2:** longitudinal projection of the dose distribution. The blue line is the dose at 1cm, while the red line is the dose on the PTV. Configurations in terms of peak width - CTC in mm are: 0.4-3.2 (1) in panel a), 0.6-1.2 (2) in panel b), 1-2 (3) in panel c), and 1-3.8 (4) in panel d).

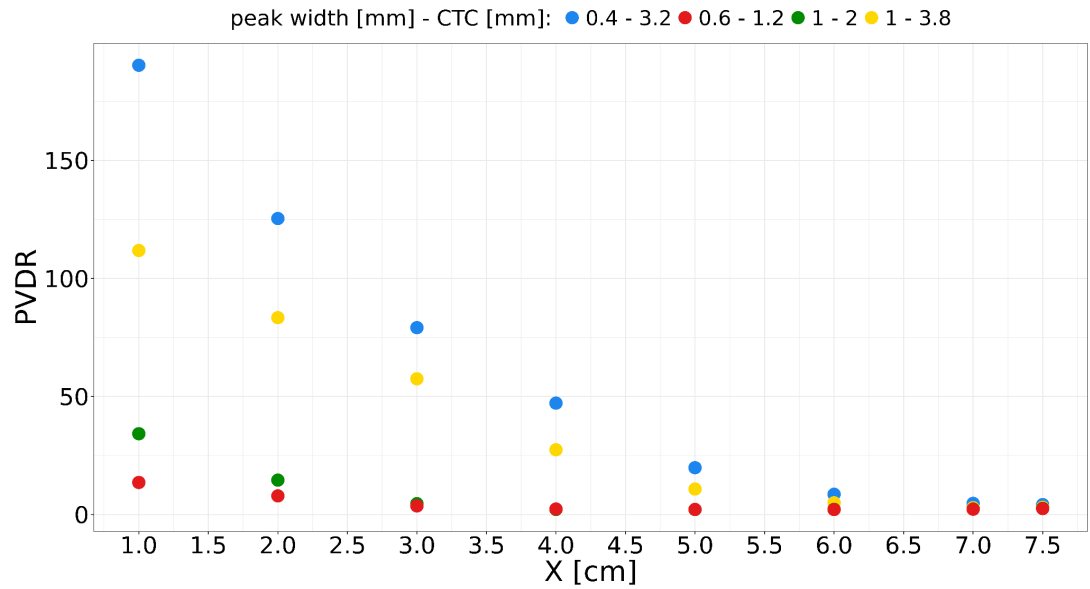

**Figure 3:** PVDR against depth in the water phantom. MB configurations in terms of peak width - CTC in mm are: 0.4-3.2 (1 - blue), 0.6-1.2 (2 - red), 1-2 (3 - green), 1-3.8 (4 - yellow).

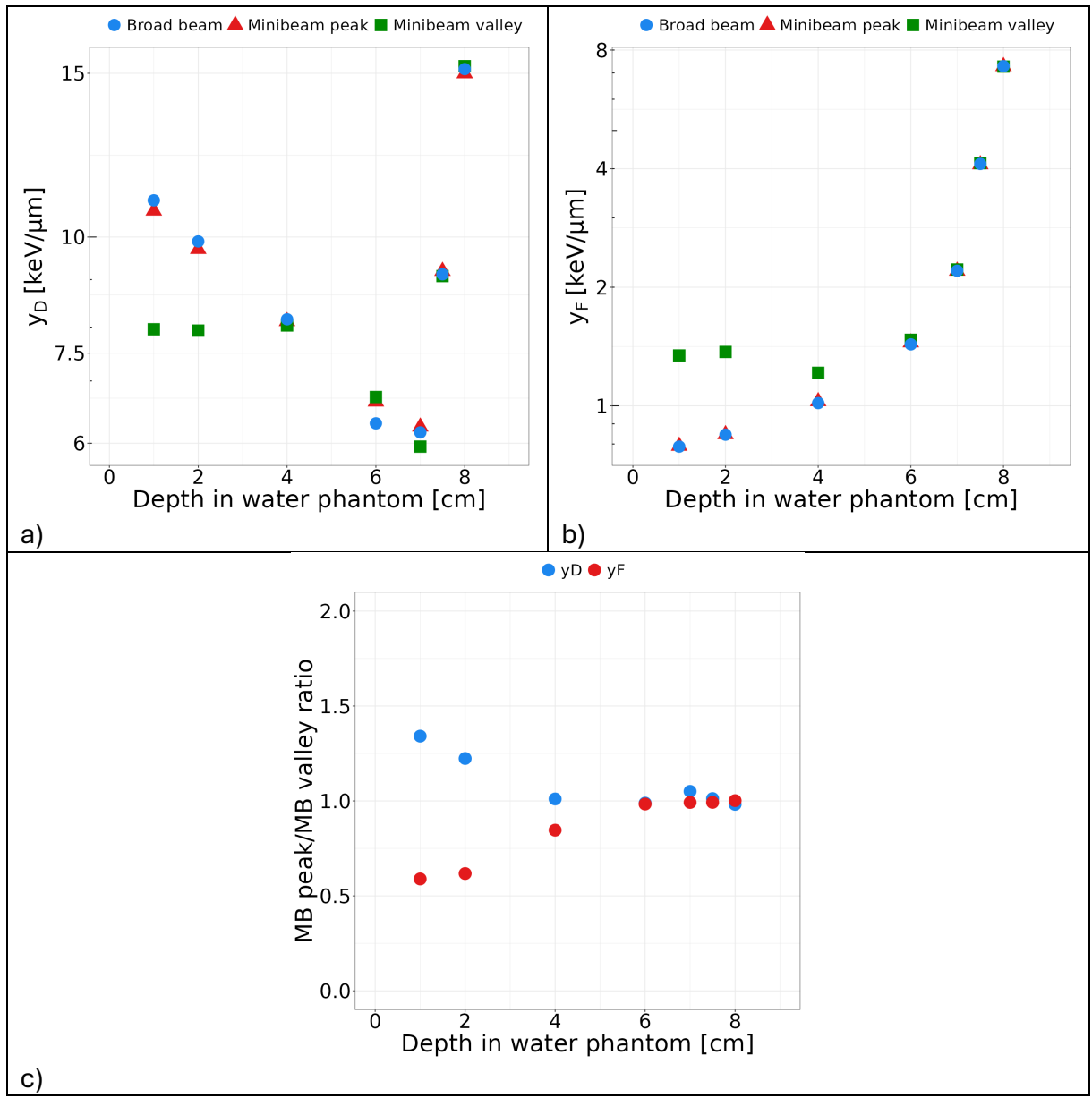

**Figure 4:** Microdosimetric estimators against depth in water phantom. Panel a) shows $y_D$ results, panel b) shows $y_F$ results, where broad beam from HF is in blue, MB peak in red and MB valley in green. On the other hand, panel c) shows the ratio between MB peak and valley for both $y_D$ (blue dots) and $y_F$ (red dots).

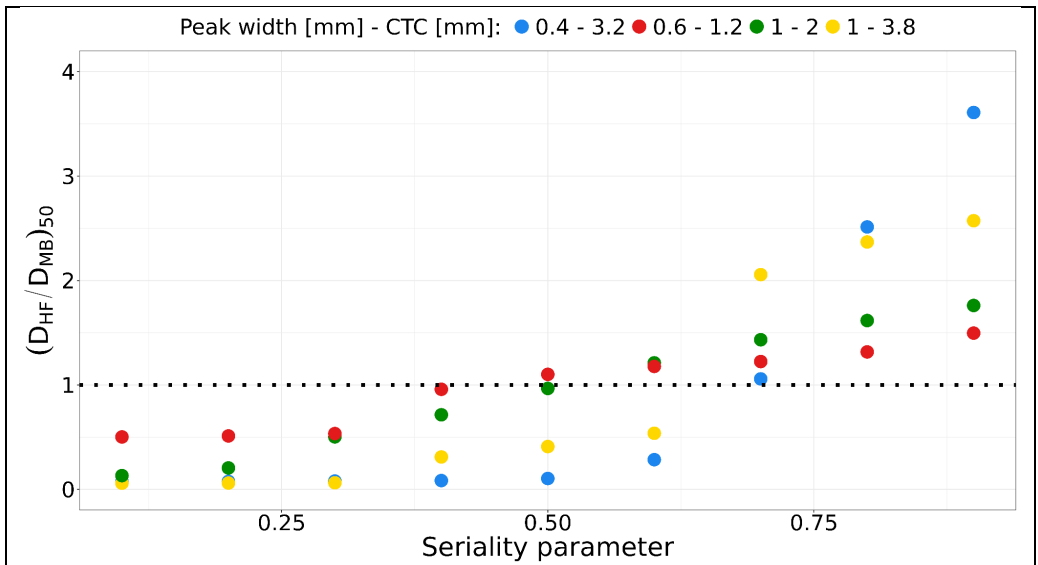
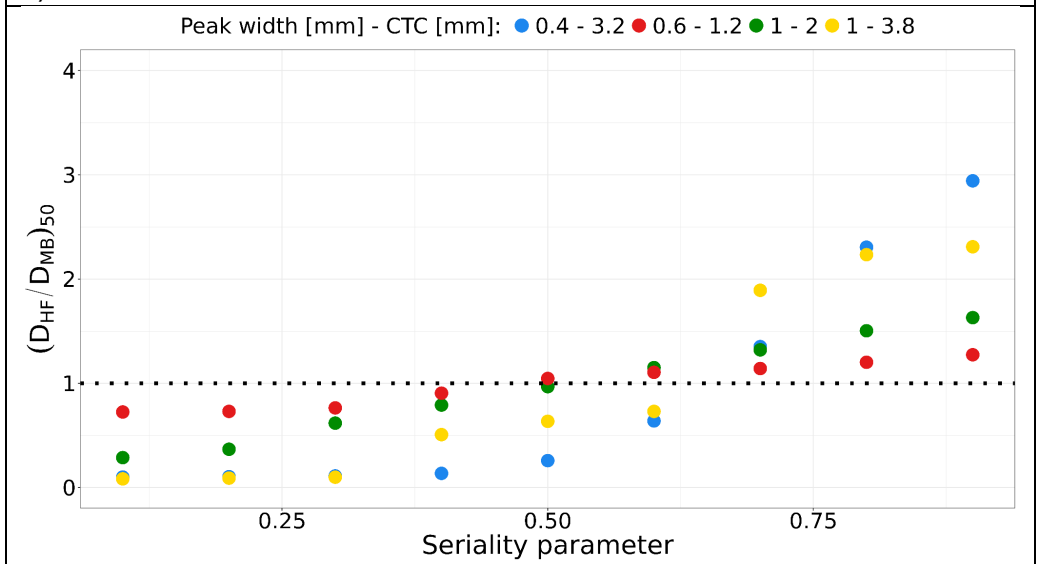
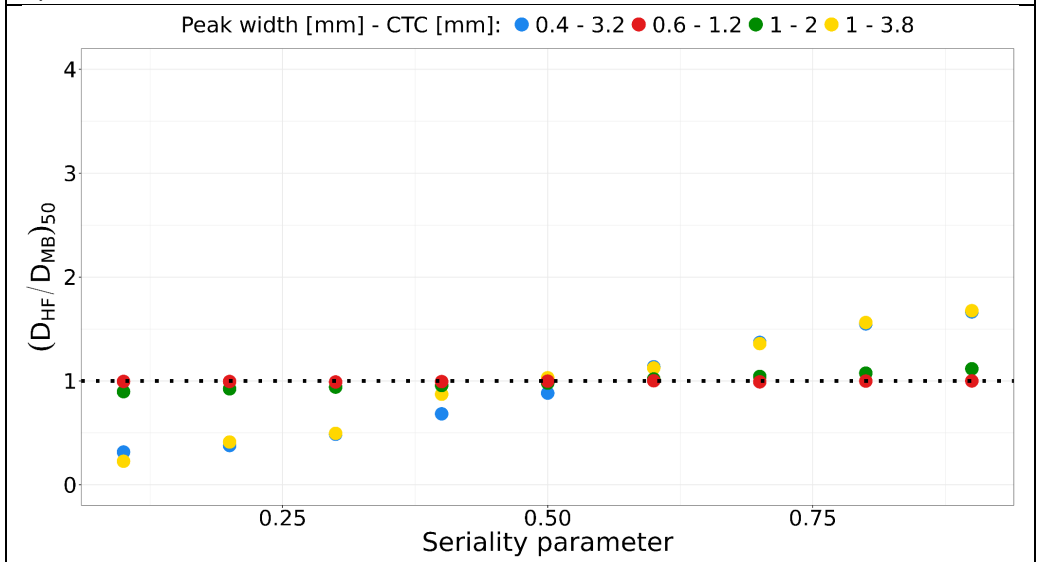

**Figure 5:** MDR$_{50}$ between HF and MB irradiation patterns against NT seriality parameter. Each panel shows results at different depth: panel a) corresponds to 1cm, panel b) 2cm and panel c) 4cm. MB configurations in terms of peak width - CTC in mm are: 0.4-3.2 (1 - blue), 0.6-1.2 (2 - red), 1-2 (3 - green), 1-3.8 (4 - yellow).

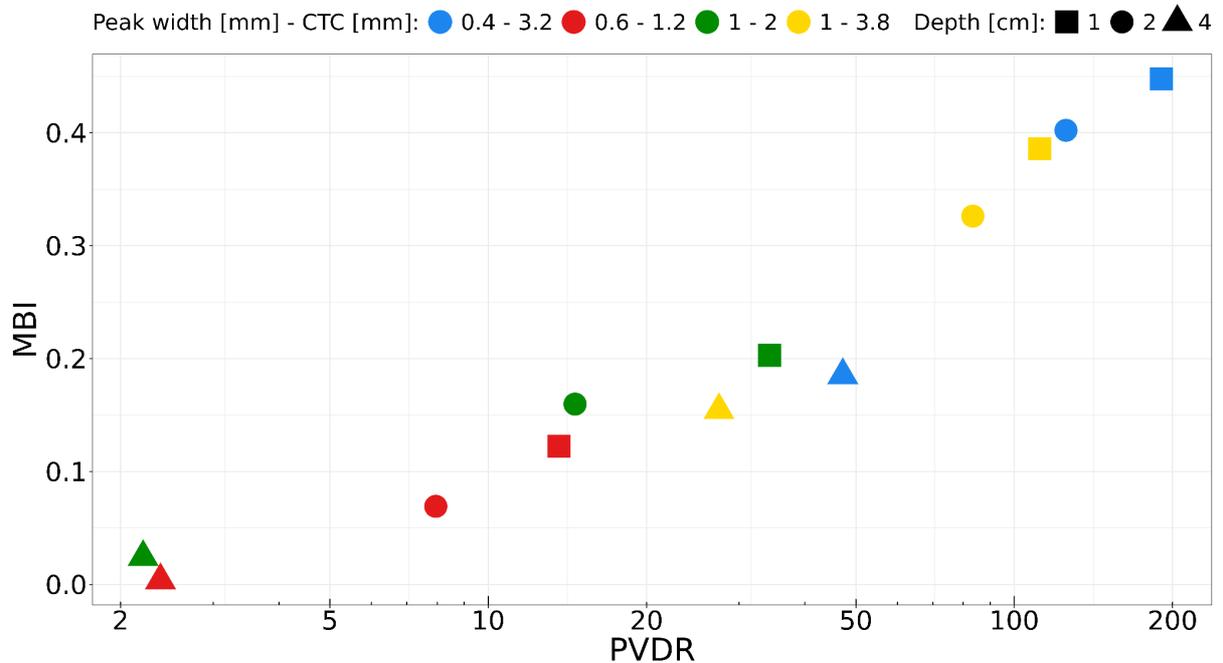

**Figure 6:** MBI against PVDR for all MB irradiation patterns at all analyzed depth values in water. MB configurations in terms of peak width - CTC in mm are: 0.4-3.2 (1 - blue), 0.6-1.2 (2 - red), 1-2 (3 - green), 1-3.8 (4 - yellow). Different shapes represent different depths: 1cm (square), 2cm (circle), 4cm (triangle).

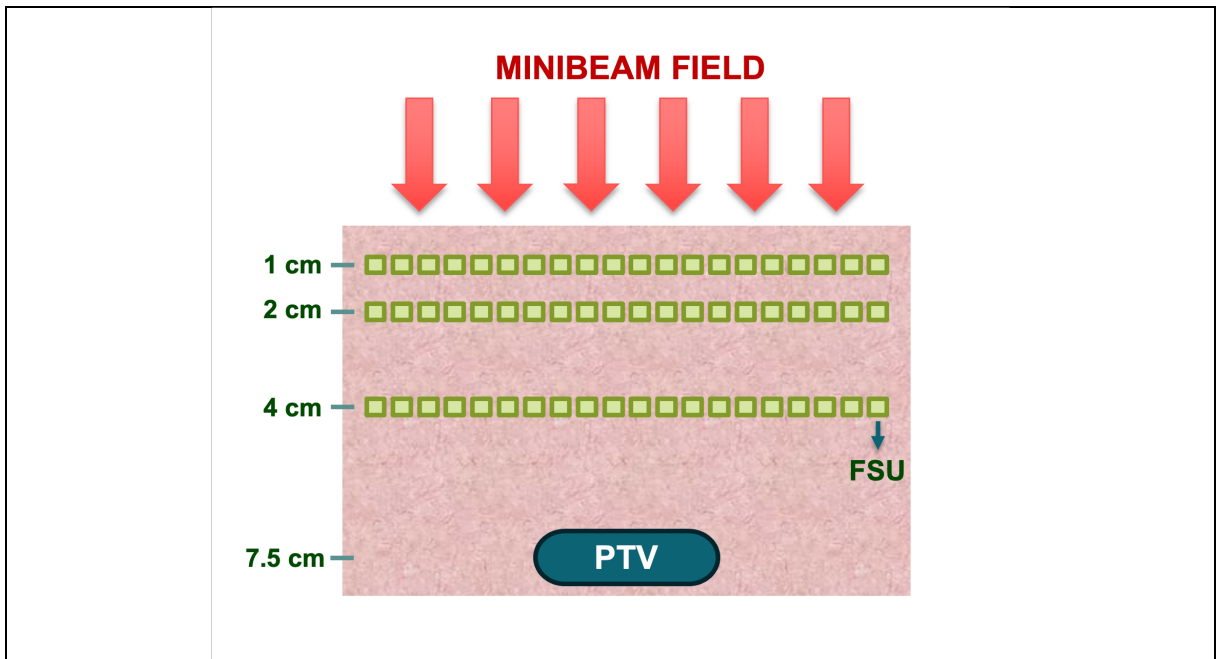

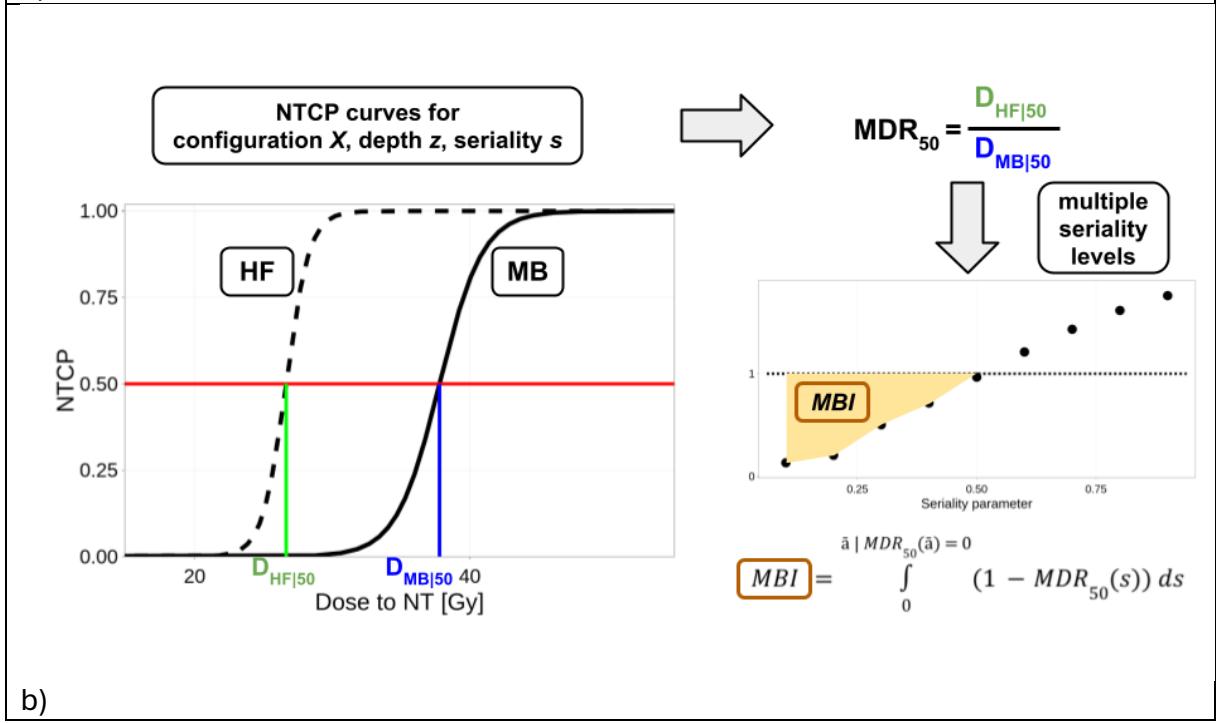

**Figure A.1**: a) MB irradiation setup considered for NTCP calculation. b) Definition scheme.

# References

[1] Yolanda Prezado et al., "On the Significance of Peak Dose in Normal Tissue Toxicity in Spatially Fractionated Radiotherapy: The Case of Proton Minibeam Radiation Therapy," *Radiotherapy and Oncology* 205 (2025): 110769.
[2] Y Prezado and Giovanna Rosa Fois, "Proton-Minibeam Radiation Therapy: A Proof of Concept," *Medical Physics* 40, no. 3 (2013): 031712.
[3] Fardous Reaz et al., "Probing the Therapeutic Window of Proton Minibeam Radiotherapy Using Dose-Response Curves in a Mouse Model," *Radiotherapy and Oncology*, 2025, 111050.
[4] Tim Schneider, "Technical Aspects of Proton Minibeam Radiation Therapy: Minibeam Generation and Delivery," *Physica Medica* 100 (2022): 64–71.
[5] Prezado et al., "On the Significance of Peak Dose in Normal Tissue Toxicity in Spatially Fractionated Radiotherapy: The Case of Proton Minibeam Radiation Therapy."
[6] Narayani Subramanian et al., "Superior Anti-Tumor Response after Microbeam and Minibeam Radiation Therapy in a Lung Cancer Mouse Model," *Cancers* 17, no. 1 (2025): 114.
[7] Annaig Bertho et al., "Evaluation of the Role of the Immune System Response After Minibeam Radiation Therapy," *International Journal of Radiation Oncology*Biology*Physics* 115, no. 2 (2023): 426–39, https://doi.org/10.1016/j.ijrobp.2022.08.011.
[8] Lorea Iturri et al., "Evaluation of Proton Minibeam Radiotherapy on Antitumor Immune Responses in a Rat Model of Glioblastoma," *Cancer Immunology Research* 13, no. 11 (2025): 1854–72, https://doi.org/10.1158/2326-6066.CIR-24-0902.
[9] Delmon Arous et al., "Integrating 2D Dosimetry and Cell Survival Analysis for Predicting Local Effect in Spatially Fractionated Radiotherapy," *Acta Oncologica* 64 (2025): 44599.
[10] Soha Bazyar et al., "Minibeam Radiotherapy with Small Animal Irradiators; in Vitro and in Vivo Feasibility Studies," *Physics in Medicine & Biology* 62, no. 23 (2017): 8924.
[11] M Dos Santos et al., "Minibeam Radiation Therapy: A Micro-and Nano-Dosimetry Monte Carlo Study," *Medical Physics* 47, no. 3 (2020): 1379–90.
[12] Darwin A Garcia et al., "Minibeam Radiation Therapy Valley Dose Determines Tolerance to Acute and Late Effects in the Mouse Oral Cavity," *International Journal of Radiation Oncology* Biology* Physics*, 2025.
[13] Stefanie Girst et al., "Proton Minibeam Radiation Therapy Reduces Side Effects in an in Vivo Mouse Ear Model," *International Journal of Radiation Oncology* Biology* Physics* 95, no. 1 (2016): 234–41.
[14] Reaz et al., "Probing the Therapeutic Window of Proton Minibeam Radiotherapy Using Dose-Response Curves in a Mouse Model."
[15] Antony Carver et al., "Design and Characterisation of a Minibeam Collimator Utilising Monte Carlo Simulation and a Clinical Linear Accelerator," *Physics in Medicine & Biology* 69, no. 13 (2024): 135001.
[16] Nathan Clements et al., "Monte Carlo Optimization of a GRID Collimator for Preclinical Megavoltage Ultra-High Dose Rate Spatially-Fractionated Radiation Therapy," *Physics in Medicine & Biology* 67, no. 18 (2022): 185001.
[17] Ludovic De Marzi et al., "Implementation of Planar Proton Minibeam Radiation Therapy Using a Pencil Beam Scanning System: A Proof of Concept Study," *Medical Physics* 45, no. 11 (2018): 5305–16.
[18] Consuelo Guardiola et al., "Optimization of the Mechanical Collimation for Minibeam Generation in Proton Minibeam Radiation Therapy," *Medical Physics* 44, no. 4 (2017): 1470–78, https://doi.org/10.1002/mp.12131.
[19] Christina Stengl et al., "Development and Characterization of a Versatile Mini-Beam Collimator for Pre-Clinical Photon Beam Irradiation," *Medical Physics* 50, no. 8 (2023): 5222–37.
[20] Guardiola et al., "Optimization of the Mechanical Collimation for Minibeam Generation in Proton Minibeam Radiation Therapy."
[21] P. Lansonneur et al., "First Proton Minibeam Radiation Therapy Treatment Plan Evaluation," *Scientific Reports* 10, no. 1 (2020): 7025, https://doi.org/10.1038/s41598-020-63975-9.
[22] Weijie Zhang et al., "Multi-collimator Proton Minibeam Radiotherapy with Joint Dose and PVDR Optimization," *Medical Physics* 52, no. 2 (2025): 1182–92, https://doi.org/10.1002/mp.17548.
[23] Emily Debrot et al., "Investigating Variable Rbe in a 12c Minibeam Field with Microdosimetry and Geant4," *Radiation Protection Dosimetry* 183, nos. 1–2 (2019): 160–66.
[24] Valentina Elettra Bellinzona et al., "Linking Microdosimetric Measurements to Biological Effectiveness in Ion Beam Therapy: A Review of Theoretical Aspects of MKM and Other Models," *Frontiers in Physics*, 2021, 623.

25 Francesco G Cordoni et al., "Cell Survival Computation via the Generalized Stochastic Microdosimetric Model (GSM2); Part I: The Theoretical Framework," *Radiation Research* 197, no. 3 (2022): 218–32.
26 Roland B Hawkins, "A Microdosimetric-Kinetic Model of Cell Killing by Irradiation from Permanently Incorporated Radionuclides," *Radiation Research* 189, no. 1 (2018): 104–16.
27 T Inaniwa and N Kanematsu, "Adaptation of Stochastic Microdosimetric Kinetic Model for Charged-Particle Therapy Treatment Planning," *Physics in Medicine & Biology* 63, no. 9 (2018): 095011.
28 M Missiaggia et al., "Cell Survival Computation via the Generalized Stochastic Microdosimetric Model (Gsm2); Part II: Numerical Results," *Radiation Research* 201, no. 2 (2024): 104–14.
29 Andrzej Niemierko and Michael Goitein, "Modeling of Normal Tissue Response to Radiation: The Critical Volume Model," *International Journal of Radiation Oncology*Biology*Physics* 25, no. 1 (1993): 135–45, https://doi.org/10.1016/0360-3016(93)90156-P.
30 Supriya Bidanta et al., "Functional Tissue Units in the Human Reference Atlas," *Nature Communications* 16, no. 1 (2025): 1526.
31 Marco Battestini et al., "Including Volume Effects in Biological Treatment Plan Optimization for Carbon Ion Therapy: Generalized Equivalent Uniform Dose-Based Objective in TRiP98," *Frontiers in Oncology* 12 (March 2022): 826414, https://doi.org/10.3389/fonc.2022.826414.
32 H. Rodney Withers et al., "Treatment Volume and Tissue Tolerance," *International Journal of Radiation Oncology*Biology*Physics* 14, no. 4 (1988): 751–59, https://doi.org/10.1016/0360-3016(88)90098-3.
33 L. A. Braby et al., "ICRU Report 98, Stochastic Nature of Radiation Interactions: Microdosimetry," *Journal of the ICRU* 23, no. 1 (2023): 1–168, https://doi.org/10.1177/14736691231211380.
34 Niemierko and Goitein, "Modeling of Normal Tissue Response to Radiation."
35 F Cordoni et al., "Generalized Stochastic Microdosimetric Model: The Main Formulation," *Physical Review E* 103, no. 1 (2021): 012412.
36 Francesco Giuseppe Cordoni, "On the Emergence of the Deviation from a Poisson Law in Stochastic Mathematical Models for Radiation-Induced DNA Damage: A System Size Expansion," *Entropy* 25, no. 9 (2023): 1322.
37 Giulio Bordieri et al., "Validation of the Generalized Stochastic Microdosimetric Model (GSM^2) over a Broad Range of LET and Particle Beam Type: A Unique Model for Accurate Description of (Therapy Relevant) Radiation Qualities," *Physics in Medicine & Biology* 70, no. 1 (2024): 015005.
38 Giulio Bordieri et al., "Integrating Nano- and Micrometer-Scale Energy Deposition Models for Mechanistic Prediction of Radiation-Induced DNA Damage and Cell Survival," *Computers in Biology and Medicine* 199 (December 2025): 111330, https://doi.org/10.1016/j.compbiomed.2025.111330.
39 P. Källman et al., "Tumour and Normal Tissue Responses to Fractionated Non-Uniform Dose Delivery," *International Journal of Radiation Biology* 62, no. 2 (1992): 249–62, https://doi.org/10.1080/09553009214552071.
40 Brita Singers Sørensen et al., "Proton FLASH: Impact of Dose Rate and Split Dose on Acute Skin Toxicity in a Murine Model," *International Journal of Radiation Oncology*Biology*Physics* 120, no. 1 (2024): 265–75, https://doi.org/10.1016/j.ijrobp.2024.04.071.
41 Marta Missiaggia et al., "Investigation of In-Field and Out-of-Field Radiation Quality With Microdosimetry and Its Impact on Relative Biological Effectiveness in Proton Therapy," *International Journal of Radiation Oncology*Biology*Physics* 115, no. 5 (2023): 1269–82, https://doi.org/10.1016/j.ijrobp.2022.11.037.
42 Bertho et al., "Evaluation of the Role of the Immune System Response After Minibeam Radiation Therapy."